\documentclass[pre,twocolumn,aps,groupaddress]{revtex4}
\usepackage[T1]{fontenc}
\usepackage[latin9]{inputenc}
\usepackage{array}
\usepackage{booktabs}
\usepackage{mathrsfs}
\usepackage{mathbbol}
\usepackage{multirow}
\usepackage{amsmath}
\usepackage{amsthm}
\usepackage{amssymb}
\usepackage{natbib}
\usepackage{stmaryrd}
\usepackage{graphicx}
\usepackage{latexsym}
\usepackage{textcomp}
\usepackage{mathtools}
\usepackage{xcolor}
\usepackage{soul}
\usepackage{adjustbox}
\usepackage[citecolor=magenta,colorlinks=true]{hyperref}
\usepackage{lineno}
\usepackage{stackrel}
\usepackage[toc,page,titletoc]{appendix}
\usepackage{bm}
\usepackage{comment}
\definecolor{mycolor1}{rgb}{0.1, 0.6, 0.6}

\begin{document}

\title{Diffusive noise controls early stages of genetic demixing}
\author{Rashmiranjan Bhutia}
\address{Tata Institute of Fundamental Research, Hyderabad, 500046, India}
\author{Stephy Jose}
\address{Tata Institute of Fundamental Research, Hyderabad, 500046, India}
\author{Prasad Perlekar}
\address{Tata Institute of Fundamental Research, Hyderabad, 500046, India}
\author{Kabir Ramola}
\address{Tata Institute of Fundamental Research, Hyderabad, 500046, India}
\date{\today}
\begin{abstract}
Theoretical descriptions of the stepping-stone model, a cornerstone of spatial population genetics, have long overlooked diffusive noise arising from migration dynamics. We derive an exact fluctuating hydrodynamic description of this model from microscopic rules, which we then use to demonstrate that diffusive noise significantly alters early-time genetic demixing, which we characterize through heterozygosity, a key measure of diversity. Combining macroscopic fluctuation theory and microscopic simulations, we demonstrate that the scaling of density fluctuations in a spatial domain displays an early-time behaviour dominated by diffusive noise. Our exact results underscore the need for additional terms in existing continuum theories and highlight the necessity of including diffusive noise in models of spatially structured populations.  
\end{abstract}
\maketitle
Understanding the mechanisms that generate and maintain genetic diversity in spatially structured populations is a fundamental problem in evolutionary biology \cite{ewens2004mathematical,hallatschek2007genetic,rana2017spreading,pigolotti2013growth,tauber2025stochastic, Murray}. Spatial population genetics, epitomized by the stepping-stone model \cite{kimura1964stepping,korolev2010genetic}, provides a framework to study allele dynamics across discrete demes connected by migration. The stepping-stone model has been instrumental in exploring how migration and genetic drift influence allele frequencies across discrete demes, and it has been very successful in quantitatively capturing fixation time (the mean time for a particular allele to become dominant) as well as genetic drift in spatially evolving populations~\cite{singha2020fixation,korolev2010genetic}.
Despite the utility of the stepping-stone model, classical theories \cite{kimura1964stepping,ewens2004mathematical} and modern continuum approaches \cite{korolev2010genetic,fisher1937wave,doering2003interacting} have focused on genetic drift and selection as well as predominantly on local demographic noise, while overlooking the stochastic fluctuations arising from migration, referred to as diffusive noise. This oversight is significant, as migration rates in natural populations can surpass birth-death rates, suggesting that diffusive noise may substantially impact genetic diversity patterns.

Coarse-grained hydrodynamic descriptions, such as the stochastic Fisher-Kolmogorov-Petrovsky-Piscounov (sFKPP) equation~\cite{doering2003interacting,korolev2010genetic,fisher1937wave,kolmogorov2019studies}, provide a continuum framework to study large-scale allele dynamics and have been often used to interpret the stepping stone model. Although these hydrodynamic models capture emergent behavior, they are often derived phenomenologically or through coarse-graining lattice models. This makes microscopic models crucial for identifying effects that may be overlooked in continuum approximations.
Recent studies \cite{duty2000broken,tauber2005applications,tauber2025stochastic,korolev2010genetic} have derived fluctuating hydrodynamic equations for similar systems but omitted noise from particle exchanges, implicitly assuming that reaction-driven fluctuations dominate. 
This gap is critical: migration rates in natural populations often exceed birth-death rates~\cite{raymond1994evolution,may1975gene,lenormand2002gene,jain1966evolutionary}, suggesting diffusive noise could profoundly influence diversity and can lead to incomplete or inaccurate representations of genetic diversity patterns. Notably, diffusive noise has also been shown to significantly modify the dynamics and phase behavior in other microscopic systems~\cite{prakash2025exact,agranov2021exact,bodineau2011phase,jose2023current, mukherjee2025hydrodynamics}.
\begin{figure}[t!]
\centering
\includegraphics[width=1\linewidth]{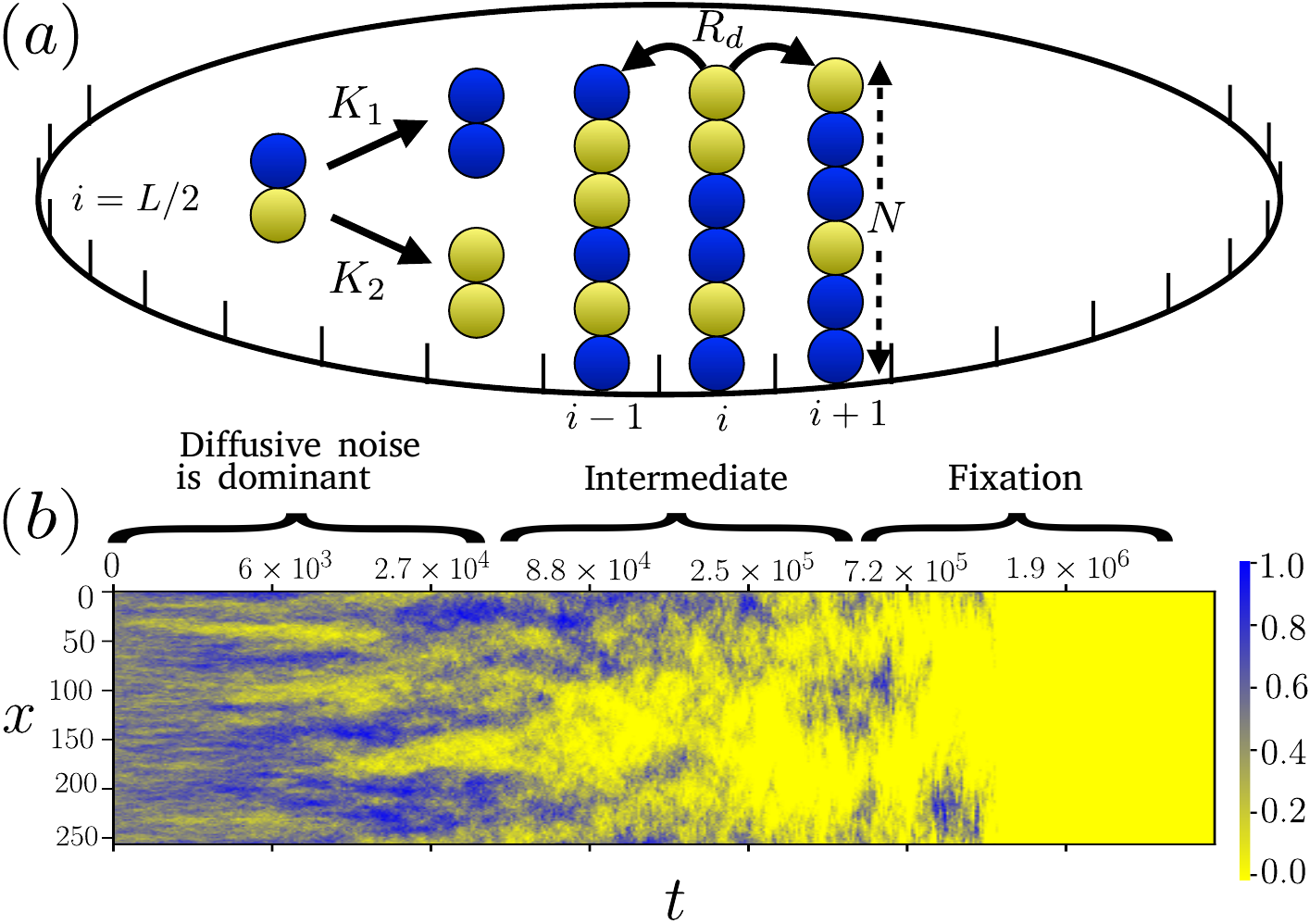}
\caption{(a) Schematic of the one-dimensional stepping-stone model. The lattice sites are indexed by $i$, with each site representing an island (or deme) inhabited by $N$ individuals of two distinct types: $A$ (blue) and $B$ (yellow). (b) Shows the evolution of the frequency $f_i = n_i^A/N$ over time, indicating three different regimes. The early-time regime features a well-mixed phase where diffusion dominates dynamics, making diffusive noise essential for computing observables. At intermediate times, reaction-driven genetic demixing forms distinct allele domains, with diffusive noise still influencing domain boundaries. The late-time fixation regime emerges when domain sizes grow to match the system size, with a single allele type present in the entire system. 
}
\label{fig: SS_Model}
\end{figure}

In this Letter, we address this gap by deriving a fluctuating hydrodynamic equation that incorporates both diffusive and reaction (demographic) noise. Our approach builds upon recent advancements in fluctuating hydrodynamics \cite{derrida2009current,bertini2015macroscopic,agranov2021exact,thompson2011lattice}, which have successfully modeled microscopic fluctuations in systems with both diffusive and reactive dynamics. We illustrate the significant effects of this diffusive noise by computing the heterozygosity as well as density fluctuations, and show how this is a much more accurate framework for understanding microscopic simulations as well as experiments.  Crucially, we show that the early time behavior of genetic demixing, quantified by heterozygosity at the origin, is well described by a modified hydrodynamic equation and not by the ubiquitously used sFKPP framework~\cite{korolev2010genetic,doering2003interacting,korolev2011competition}.

{\it Stepping Stone Model:} We consider a one-dimensional stepping-stone model (see Fig.~\ref{fig: SS_Model}) consisting of a periodic lattice of length $ l = La $, with lattice sites indexed by $ i = -L/2+1, ..., -1, 0, 1, ..., L/2 $ and separated by a distance $ a $. Each site represents a deme that contains a fixed number $N$  of individuals belonging to two allelic types, denoted $ A $ and $ B$. The system evolves through local reproduction and short-range migration processes, which are modeled using discrete-time update rules.

At each time step, the following processes can occur:
(1) Reproduction: 
Two individuals are randomly selected within a deme, and one of them reproduces while the other is removed, maintaining the population size $N$ at each site fixed. Reproduction is modeled by the following reactions~\cite{Moran}:
\begin{equation}
 A + B \xrightarrow{K_1} A + A, \quad A + B \xrightarrow{K_2} B + B,
 \label{reaction_rates}
   \end{equation}
   where $ K_1 $ and $ K_2 $ are the rates of reproduction of alleles $ A $ and $ B $, respectively.
(2) Migration: With rate $ R_d $, an individual at site $i$ is exchanged with an individual from a neighboring site $ i+1 $. This model includes stochastic effects due to genetic drift, leading to fixation dynamics and spatial heterogeneity in allele frequencies. The key observable of interest is the fraction $ f_i(\tau_j) $ of allele $ A $ at site $ i $ at time step $ \tau_j $, which evolves under the combined influence of reactions and diffusion. This is defined as $f_i(\tau_j) = n_i^A/N$, where $n_i^A$ is the number of $A$ alleles. The fraction of allele $B$ is then simply given by $1-f_i(\tau_j)$, since the total number of individuals per site is conserved. Our simulations employ the same microscopic rules, whose details as well as additional explanations related to the figures in the main text are provided in the Supplemental Material~\cite{SI}.

{\it Derivation of Fluctuating Hydrodynamics:} To derive the fluctuating hydrodynamics of the stepping-stone model, we start from the discrete description and extend it to a continuum framework \cite{lefevre2007dynamics,doi1976second,peliti1985path}. The change in allele frequency at site $i$ during a microscopic time step $d\tau$ is captured by the quantity $J_i(\tau_j) = f_i(\tau_{j+1}) - f_i(\tau_j)$, which we refer to as the microscopic current. The probability of observing a specific trajectory of frequencies $\{f_i(\tau_j)\}$ can be expressed in terms of these currents as  
\begin{equation}
    P(\{f\}) = \left\langle \prod_{i,j} \delta\left(f_i(\tau_{j+1}) - f_i(\tau_j) - J_i(\tau_j)\right) \right\rangle_{\{J\}},
\end{equation}
where the delta functions enforce the dynamical rules, and the average is taken over all possible realizations (or histories) of $\{ J \}$. Next, using the integral representation of the delta function, $\delta(a-b) = \int_{-i\infty}^{i\infty} \frac{d\hat{f}}{2\pi i} e^{-\hat{f}(a-b)}$, the probability of observing a trajectory is rewritten as a path integral over auxiliary fields $\hat{f}_i(\tau_j)$ as
\begin{equation}
    P(\{f\}) = \int \mathcal{D}\hat{f} \, e^{-S}.
\end{equation}
Here, \( \mathcal{D}\hat{f} = \prod_{i=-L/2+1}^{L/2} \prod_{j=1}^{M} d\hat{f}_i(\tau_j) \) denotes the path integral measure over the auxiliary fields at each site \( i \) and time step \( \tau_j \) and the action $S$ is given by
\begin{eqnarray}
\begin{split}
     S = \sum_{i,j} \left[ \hat{f}_i(\tau_j) \left(f_i(\tau_{j+1}) - f_i(\tau_j)\right) \right] \\-
 \underbrace{\ln \left[\prod_j\left\langle e^{\sum_i\hat{f}_i(\tau_j) J_i(\tau_j)}\right\rangle \right]}_{(T_d+T_r)}.
\label{eq:Action_general}
\end{split}
\end{eqnarray}
Eq.~\eqref{eq:Action_general} holds for any process governed by a microscopic stochastic rule, whereas the specific form of the second term depends on the underlying process. This path-integral approach provides a systematic framework to derive the continuum limit. Given that the probability of each event is known, the calculation of the average in Eq.~\eqref{eq:Action_general} becomes straightforward~\cite{martin2021nonequilibrium,mukherjee2025hydrodynamics}.
The second term of the above action has been divided into contributions from two sources, $T_d$ and $T_r$, which can be derived from the microscopic rules of the stepping stone model.  The diffusive part \(T_d\) which arises from particle exchanges between neighboring sites \(i\) and \(i+1\) is given as
\begin{eqnarray}
 T_d &=& R_d d\tau \sum_{i,j} \Big[ f_i(\tau_j)(1-f_{i+1}(\tau_j)) \left(e^{\frac{\hat{f}_{i+1}(\tau_j) - \hat{f}_i(\tau_j)}{N}} - 1\right) \nonumber\\&+& f_{i+1}(\tau_j)(1-f_i(\tau_j)) \left(e^{\frac{\hat{f}_i(\tau_j) - \hat{f}_{i+1}(\tau_j)}{N}} - 1\right) \Big]. 
 \label{eq:T_d}
\end{eqnarray}
A similar expression for $T_d$ has been reported for the asymmetric simple exclusion process (ASEP)~\cite{derrida1998exactly,spohn2012large} as well as for other related models~\cite{martin2021nonequilibrium,thompson2011lattice,mukherjee2025hydrodynamics}.
The reaction term \(T_r\), which captures birth-death processes within a deme, is given as 
\begin{eqnarray}
    T_r& =& d\tau \sum_{i,j} \Big[ K_1 f_i(\tau_j)(1-f_i(\tau_j)) \left(e^{\frac{\hat{f}_i(\tau_j)}{N}} - 1\right) \nonumber\\&+&
    K_2 f_i(\tau_j)(1-f_i(\tau_j)) \left(e^{-\frac{\hat{f}_i(\tau_j)}{N}} - 1\right) \Big].
    \label{eq:T_r}
\end{eqnarray}
We next use this form of the action to derive the continuum fluctuating hydrodynamic equation for the stepping-stone model. To do so, we perform a coarse-graining procedure. The spatial and temporal coordinates are scaled as $x = ia$ and $t = jd\tau$, with $a, d\tau \to 0$.
To ensure hydrodynamic scaling, the exchange rate $R_d$ is rescaled as $R_d \to \tilde{R}_d/a^2$, ensuring that diffusive noise and reaction noise contribute equally in the hydrodynamic limit. 
We expand the discrete fields $f_{i\pm1}(\tau_j)$ and $\hat{f}_{i\pm1}(\tau_j)$ in Taylor series: $f(x \pm a, t) \approx f(x,t) \pm a \partial_x f + \frac{a^2}{2} \partial_x^2 f$, and 
$\hat{f}(x \pm a, t) \approx \hat{f}(x,t) \pm a \partial_x \hat{f} + \frac{a^2}{2} \partial_x^2 \hat{f}$, and substitute into the action $S$. Here, $ f(x,t) $ is the continuous representation of the allele fraction. Retaining terms up to $\mathcal{O}(a^2, d\tau)$, we find that the action separates into contributions from diffusion and reactions as:
\begin{align}
\begin{split}
S = \frac{1}{a} \int dx\,dt \big[ \hat{f} \partial_t f + D_d \partial_x f \partial_x \hat{f} -\frac{\sigma_s}{2} \hat{f}\\- \frac{\sigma_d}{2} (\partial_x \hat{f})^2 - \frac{\sigma_r}{2} \hat{f}^2\big],
\end{split}
\end{align}
where $D_d = \tilde{R}_d /N$ is the macroscopic diffusion constant, while $\sigma_d = 2D_d  f(1-f)/N$, $\sigma_r = 2D_r  f(1-f)/N$ and $\sigma_s = 2S f(1-f)/N$ are noise amplitudes associated with diffusion, reaction and selection (deterministic drift) respectively. In addition, $D_r = R_r/N$, $R_r = (K_1 + K_2)/2$, and $S = K_1 - K_2$ represent the genetic diffusion constant, neutral reaction rate, and selective advantage, respectively.  
Next, following standard techniques \cite{lefevre2007dynamics,pechenik1999interfacial,martin1973statistical,derrida2005fluctuations}, we employ a Hubbard-Stratonovich transformation so that the quadratic terms in $\hat{f}$ are linearized by introducing Gaussian noise fields $\eta_d(x,t)$ (conservative) for the diffusive term $\frac{\sigma_d}{2}(\partial_x \hat{f})^2$ and $\eta_r(x,t)$ (non-conservative) for the reaction term $\frac{\sigma_r}{2}\hat{f}^2$. This yields the probability of observing a history of allele frequencies as
\begin{equation}
   P(\{f\}) = \int \mathcal{D}\hat{f}  D \eta_d D \eta_r \, e^{-S}. 
\end{equation}
where, \(\int \mathcal{D}\hat{f} D\eta_d D\eta_r\) denotes the path integral measure over the auxiliary fields \(\hat{f}(x,t)\), and the noise fields \(\eta_d(x,t)\) and \(\eta_r(x,t)\), at each space-time point and
\begin{eqnarray}
S &=& \frac{1}{a} \int dx\,dt \big[ \hat{f} \partial_t f -\hat{f} D_d \partial_x^2 f- \hat{f} \partial_x\left(\sqrt{\sigma_d }\eta_d\right)  \nonumber\\&-&  \hat{f}\frac{\sigma_s}{2} - \hat{f} \sqrt{\sigma_r }\eta_r +\frac{\eta_d^2}{2}+\frac{\eta_r^2}{2} .
\label{hydro_action}
\end{eqnarray}

Performing an integral over $\hat f$, we obtain
\begin{small}
\begin{eqnarray}
 P\left(\{f\}\right) &\approx& \int  D\eta_d D\eta_r~e^{-\frac{1}{2a}\int dxdt \left( {\eta_d^2}+{\eta_r^2}\right)}\times \nonumber\\&&\delta\Bigg(\partial_t f-D_d \partial_x^2 f - \frac{\sigma_s}{2} - \partial_x\left(\sqrt{\sigma_d a}\eta_d\right) - \sqrt{\sigma_r a}\eta_r \Bigg). 
 \nonumber\\
 \label{path_integral_cont2}
\end{eqnarray}
\end{small}
This leads to the stochastic partial differential equation: 
\begin{equation}
\partial_t f = D_d \partial_x^2 f + \frac{S}{N} f(1-f) + \partial_x\left(\sqrt{\sigma_d a}\eta_d\right) + \sqrt{\sigma_r a}\eta_r,
\label{sde_allele}
\end{equation}
where $\eta_d$ and $\eta_r$ are Gaussian white noises with mean zero and correlations given by
\begin{eqnarray}
\langle\eta_d (x_1,t_1)\eta_d (x_2,t_2)\rangle&=&\delta (x_1-x_2)\delta (t_1-t_2),\nonumber\\
\langle\eta_r (x_1,t_1)\eta_r (x_2,t_2)\rangle&=&\delta (x_1-x_2)\delta (t_1-t_2),
\nonumber\\
\langle\eta_d (x_1,t_1)\eta_r (x_2,t_2)\rangle&=&0.
\label{correlations2}
\end{eqnarray} 
We note that Eq.~\eqref{sde_allele} without the presence of the third term is popularly known as the sFKPP equation~\cite{doering2003interacting,korolev2010genetic,fisher1937wave,kolmogorov2019studies}, which is a common model to describe the invasion of advantageous organisms within an existing population. To highlight the effects of diffusive noise, we focus on the case where the selective advantage is zero, i.e., we set the second term in Eq.~\eqref{sde_allele} to zero. This allows us to isolate and study the impact of stochastic effects due to migration and genetic drift in detail.

\begin{figure}
    \centering
  \includegraphics[width=1\linewidth]{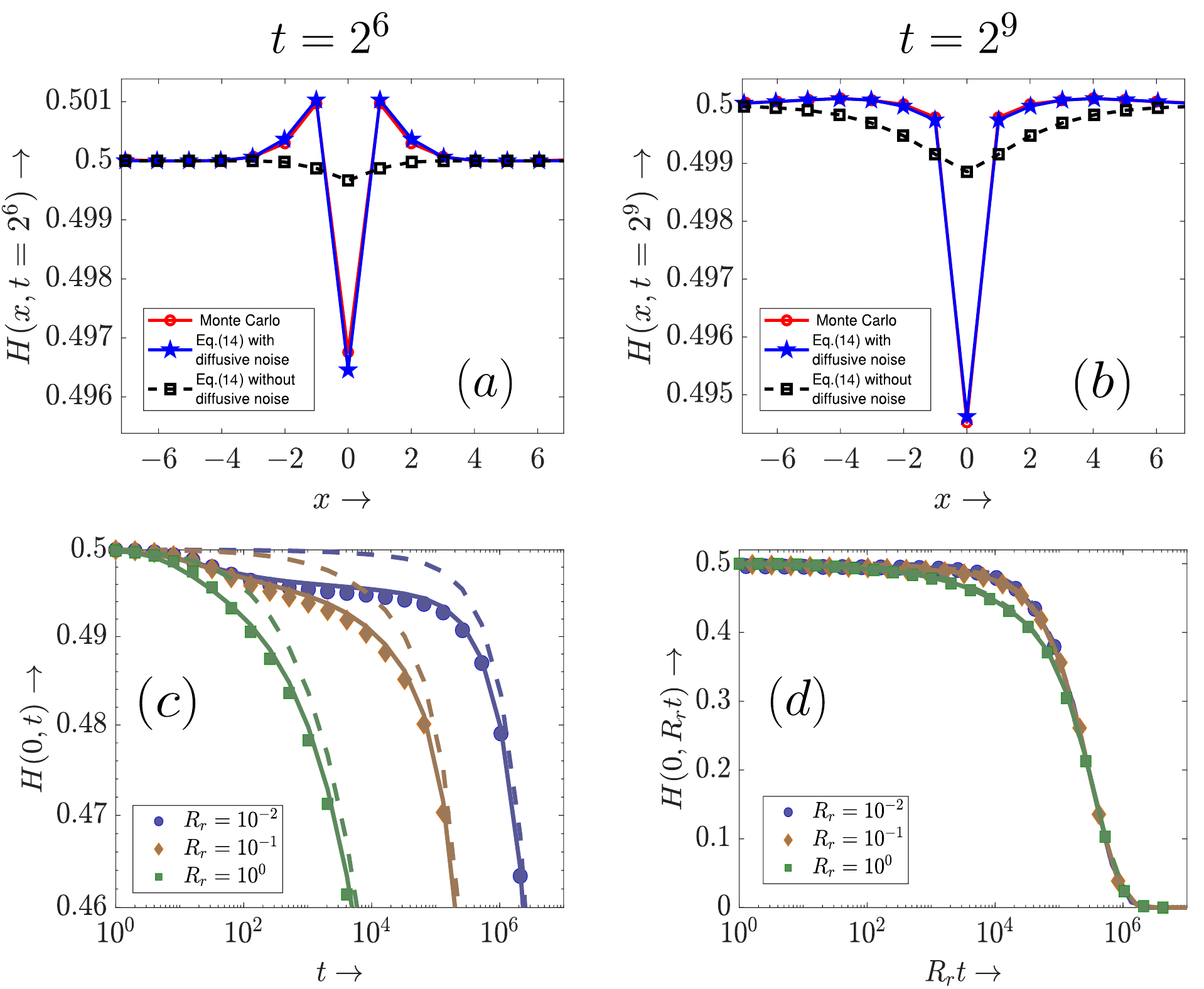}   \caption{Effect of diffusive noise on spatial heterozygosity $H(x,t)$: 
  (a) and (b) display $H(x,t)$ at different time instants. The red circles represent data from direct Monte Carlo simulations of the stepping stone model; stars indicate solutions of Eq.~\eqref{eq:Hetero_new} that include diffusive noise, and squares correspond to solutions without this noise.  (c) illustrates the time evolution of heterozygosity at the origin $H(0,t)$ for various reaction rates. The points are obtained from Monte Carlo simulations. The solid and dashed lines represent solutions of Eq.~\eqref{eq:Hetero_new} with and without diffusive noise, respectively. (d) presents the long-time decay of heterozygosity for different reaction rates, with the time axis rescaled by the reaction rate. All curves display the same asymptotic behavior at large times~\cite{SI}. 
 }
\label{fig:hxt}
\end{figure}

{\it Effect on Heterozygosity:}
To illustrate the consequences of the diffusive noise, we focus on the two-point correlation function $ H(x_1, x_2, t) = \langle f(x_1, t)(1 - f(x_2, t)) + f(x_2, t)(1 - f(x_1, t)) \rangle $, commonly referred to as the average spatial heterozygosity in population genetics. This represents the average probability that two individuals at positions $ x_1 $ and $ x_2 $ carry different alleles and is routinely used to quantify genetic diversity \cite{kimura1964stepping,korolev2010genetic,korolev2011competition}.
Starting from the fluctuating hydrodynamic equation for the frequency of the allele given in Eq.~\eqref{sde_allele}, we derive the equation for $ H $. Following previous work \cite{korolev2010genetic}, 
we apply Ito's Lemma, and expand $H$ to second order in $ \delta f $, using the stochastic increments $ \delta f(x, t) = D_d \partial_x^2 f \, \delta t + dw(x, t) $, where $ dw $ encodes the noise correlations, we have
\begin{eqnarray}
\nonumber
\langle dw(x, t) dw(x', t') \rangle = \left( \frac{2D_d a}{N} \partial_x \partial_{x'} [f(1 - f)\delta(x - x')]  \right.\\
\left.+ \frac{2D_r a}{N} f(1 - f)\delta(x - x') \right) \delta(t - t').
\nonumber\\
\end{eqnarray}
An analogous form of averaging arises in general Langevin processes describing systems with diffusive dynamics~\cite{dean1996langevin}. After averaging over the noise, the deterministic terms yield $ D_d (\partial_{x_1}^2 + \partial_{x_2}^2)H $, while the noise contributions produce:  
$- \frac{2D_r a}{N} H \delta(x_1 - x_2)$ and $ + \frac{2D_d a}{N} \partial_{x_1} \partial_{x_2}[H \delta(x_1 - x_2)]$ respectively.
The first term reflects reaction-driven diversity loss as seen in previous studies \cite{korolev2010genetic,korolev2011competition,pigolotti2013growth}, while the second term represents a new contribution arising from diffusive noise. The resulting equation for the heterozygosity $H(x_1,x_2)$ becomes
\begin{equation}
\partial_t H = 2D_d \partial_x^2 H - \frac{2D_r a}{N} H \delta(x) + \frac{2D_d a}{N} \partial_x^2 [H \delta(x)],
\label{eq:Hetero_new}
\end{equation}
where $ x = x_1 - x_2 $. This generalizes previous equations that did not include diffusive noise \cite{korolev2010genetic, brunet1997shift}.
Our modified equation introduces an additional term, the third term arising uniquely from diffusive noise, which modifies the spatial structure of heterozygosity. Specifically, it suppresses diversity at the origin while enhancing it at other spatial locations at short times. This behavior is illustrated by the heterozygosity profiles shown in Fig.~\ref{fig:hxt}. We observe a clear discrepancy between Monte Carlo simulations and the theory without diffusive noise. 
However, when diffusive noise is correctly accounted for, the simulations match the theory perfectly. Previous theories without diffusive noise do not capture this, leading to discrepancies in spatial profiles when $ R_r \ll R_d $.
This analysis shows that early genetic demixing is accurately described by a hydrodynamic equation with diffusive noise rather than the sFKPP equation. However, the long-time behavior remains the same regardless of the presence of diffusive noise (see Supplemental Material for details \cite{SI}).

{\it Current Fluctuations:}
To further demonstrate the importance of diffusive noise, we next focus on an additional quantity, namely the fluctuations in the number of alleles across a spatial domain.
The total flux of particles $Q_T$ across the origin up to a time $T$ has become a standard observable in diffusive systems, both in and out of equilibrium, over the past few decades~~\cite{derrida2009current2,derrida2009current,krapivsky2012fluctuations,banerjee2022role,krapivsky2015tagged,krapivsky2014large,mallick2022exact,dandekar2022dynamical,jose2023current,jose2023generalized}. It is most commonly used as a test of large deviations using macroscopic fluctuation theory (MFT)~\cite{derrida2005fluctuations,bertini2005current,bertini2006non,bertini2007stochastic,bertini2009towards,krapivsky2012fluctuations,dandekar2024current}. In the following, we use this quantity to determine the relative contribution of diffusive noise in the stepping-stone model.
We define the net change in allele concentration $ A $ up to time $T$ in half the spatial domain (which includes contributions from both flux and reactions), also called the integrated current:
\begin{equation}
 Q_T = \frac{1}{a} \int_0^{l/2} [f(x, T) - f(x, 0)] \, dx.    
\end{equation}
The variance of $ Q_T $, denoted $ \langle Q_T^2 \rangle_c $, provides insight into the fluctuations in allele transport within the system. To compute this variance, we employ the MFT framework, which facilitates the analysis of fluctuations in systems governed by hydrodynamic equations \cite{bertini2005current}.
Within the MFT framework, the probability of observing a specific trajectory of the system is characterized by an action functional $ \mathcal{S}$. This action $\mathcal{S}$ is the same as Eq.~\eqref{hydro_action} but now biased by an extra term $\lambda q_T$, i.e., $\mathcal{S}=S-\lambda q_T$. The moment generating function for $ Q_T $, $ \langle e^{\lambda Q_T} \rangle $, can be expressed as a path integral:
\begin{equation}
\langle e^{\lambda Q_T} \rangle = \int \mathcal{D}f \, \mathcal{D}\hat{f} \, e^{-L\mathcal{S} },
\end{equation}
where $ \hat{f} $ is the conjugate field, $ q_T = \int_0^{1/2} [f(x', T) - f(x', 0)] \, dx' $ represents the scaled integrated current. To evaluate this path integral, we apply a saddle-point approximation~\cite{krapivsky2012fluctuations}, leading to a set of coupled Euler-Lagrange equations for the fields $ f(x, t) $ and $ \hat{f}(x, t) $:
\begin{eqnarray}
\begin{aligned}
\partial_t f &= D_d \partial_x^2 f - \partial_x \left( \sigma_d(f) \partial_x \hat{f} \right) + \sigma_r(f) \hat{f}, \\
\partial_t \hat{f} &= -D_d \partial_x^2 \hat{f} - \frac{\sigma_d'(f)}{2} (\partial_x \hat{f})^2 - \frac{\sigma_r'(f)}{2} \hat{f}^2.
\end{aligned}
\end{eqnarray}
Here, the primes denote derivatives with respect to $ f $.

Next, we apply a standard perturbative expansion to compute the current fluctuations \cite{krapivsky2012fluctuations,jose2023current}.
We expand both fields in powers of the perturbation parameter $ \lambda $, which quantifies deviations from the typical trajectory:  
\begin{equation}
f = f_0 + \lambda f_1 + \lambda^2 f_2 + \cdots, \quad \hat{f} = \lambda \hat{f}_1 + \lambda^2 \hat{f}_2 + \cdots.
\end{equation}
At linear order in $ \lambda$, the Euler-Lagrange equations become 
\begin{equation}
\begin{aligned}
\partial_t f_1 &= D_d \partial_x^2 f_1 - \partial_x \left( \sigma_d(f_0) \partial_x \hat{f}_1 \right) + \sigma_r(f_0) \hat{f}_1, \\
\partial_t \hat{f}_1 &= -D_d \partial_x^2 \hat{f}_1.
\label{eq_linearized_EL}
\end{aligned}
\end{equation}
The boundary conditions are $f_1(x,0) = 0 $ (quenched initial state) and $ \hat{f}_1(x,T) = \theta(x) $~\cite{derrida2009current,dandekar2024current,jose2023current}, where $ \theta(x) = 1 $ for $ x \in [0, 1/2] $. The homogeneous solution for $ \hat{f}_1 $, backward in time ($ \tau = T - t $) can be easily found in Fourier space to be
\begin{equation}
\hat{f}_1(k, \tau) = \frac{i(1 - e^{ik\pi})}{2k\pi} e^{-D_d(2\pi k)^2 \tau}.
\label{eq_f1_fourier}
\end{equation} 
\begin{figure}[t!]
\centering
\hspace{-1cm}
\includegraphics[width=1\linewidth]{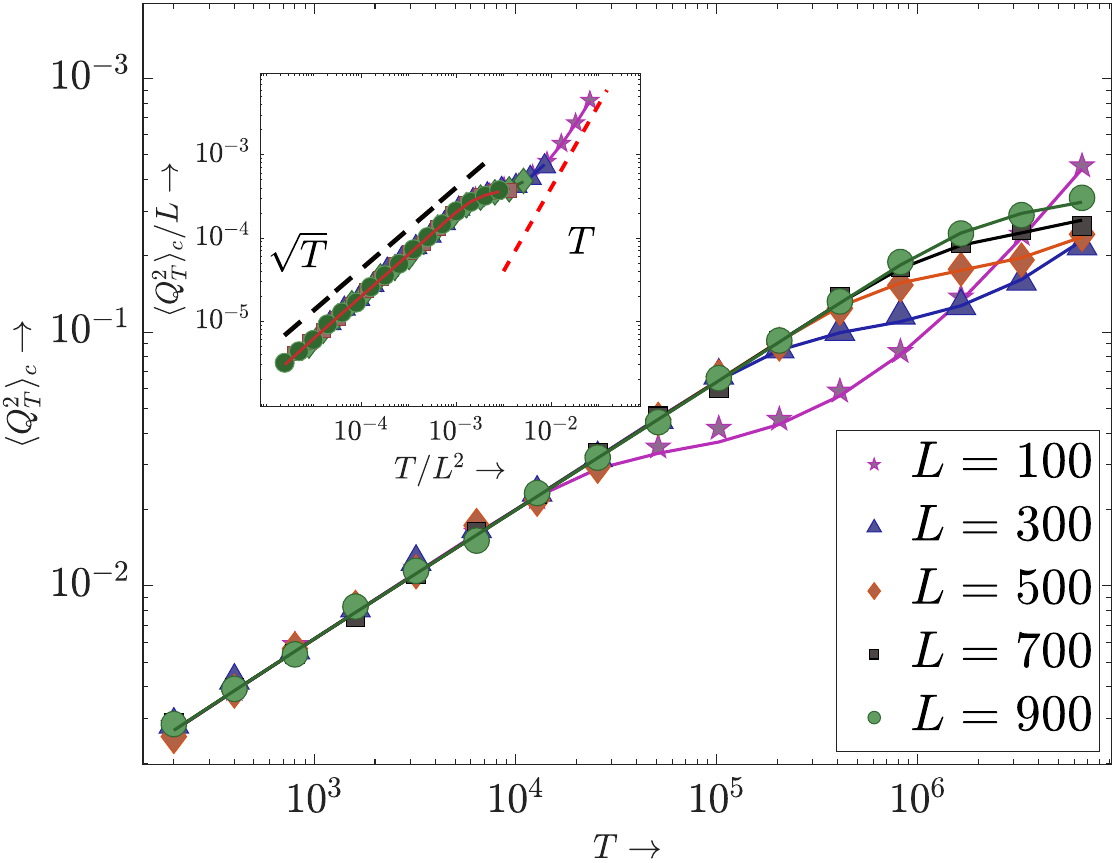}
\caption{The variance of the current $Q_T$ plotted against time for different values of $L$. The points represent the results obtained from Monte Carlo simulations, and the dashed lines are obtained from the MFT prediction (performing the summation in Eq.~\eqref{eq:Q2Tk}). 
{\bf (Inset)} The variance is plotted in rescaled coordinates, where the temporal coordinate is scaled as $T/L^2$ and the variance is scaled as ${\langle{Q_T}^2\rangle}_c/L$. All the curves for different system sizes collapse onto a single curve.  For $T \ll T^*$, $\langle Q_T^2 \rangle \sim \sqrt{T}$ (diffusive noise dominates); whereas for $T \gg T^*$, $\langle Q_T^2 \rangle \sim T$ (reaction noise), confirming the theoretical scaling.  
}
\label{fig:Q2_Dg}
\end{figure}

At leading order ($ \lambda^2 $), the variance $ \langle Q_T^2 \rangle_c $ becomes:  
\begin{equation}
\langle Q_T^2 \rangle_c = \sum_{k=-\infty}^\infty \int_0^T dt \left[ \sigma_d k^2 |\hat{f}_1(k, t)|^2 + \sigma_r |\hat{f}_1(k, t)|^2 \right],
\label{eq:Q2Tk}
\end{equation} where $ \hat{f}_1(k, t) $ given in Eq.~\eqref{eq_f1_fourier} solves the linearized MFT equations in Eq.~\eqref{eq_linearized_EL}. The predictions from the above expression are compared with the results from direct numerical simulations of the stepping-stone model in Fig.~\ref{fig:Q2_Dg}, showing perfect agreement.
It is straightforward to analyze the scaling limits of the above expression: at short times ($ T \ll T^* $) diffusive noise dominates and the sum converges to a scaling form $\langle Q_T^2 \rangle_c \sim {\sigma_d}\sqrt{T}/{\sqrt{2\pi D_d}} $, while at long times ($ T \gg T^* $) reaction noise prevails, yielding $
\langle Q_T^2 \rangle_c \sim {\sigma_r T}/{8} $, 
with a crossover timescale given by $ T^* \sim 32\sigma_d^2 / (\pi\sigma_r^2 D_d) $. The $\sqrt{T}$ scaling is typical of diffusive systems~\cite{derrida2009current2,krapivsky2012fluctuations} while the linear $T$ behavior is typical of reactive as well as active systems~\cite{banerjee2020current,jose2023generalized}.
Interestingly, the short-time behavior of current fluctuations in the stepping stone model is the same as the symmetric simple exclusion process (SSEP) \cite{levitt1973dynamics,arratia1983motion,derrida2005fluctuations, krapivsky2012fluctuations,spohn2012large}, since in the absence of reaction dynamics the exchange rules are similar, however the fluctuations are reduced by a factor of $1/N$ compared to the SSEP.

{\it Conclusions:} 
Our analysis shows that both types of noise are essential for accurate predictions in spatially structured populations, particularly at short times when migration rates exceed reaction rates ($R_d \gg R_r$). This insight extends beyond population genetics to general reaction-diffusion systems where transport and local reactions coexist, such as epidemic spreading, ecological invasions, or chemical front propagation~\cite{noble1974geographic,campos2008lattice,schimansky1991kink}.  While the classical stochastic Fisher equation~\cite{doering2003interacting,korolev2011competition,pigolotti2013growth} and related continuum models \cite{hallatschek2009fisher,korolev2010genetic} focus solely on reaction noise (genetic drift), they miss initial fluctuations when migration prevails.
Our exact results demonstrate that diffusive noise is crucial for correctly modeling early-time genetic demixing in populations. 
We have provided quantifiable observables such as heterozygosity as well as current fluctuations, which offer a direct measure to identify the timescales where diffusive noise is relevant. This limitation is crucial considering experiments, where high migration rates in microbial expansions cause rapid allele mixing and reduce local heterozygosity. We therefore hope our study motivates future experimental and numerical studies designed to investigate early-time genetic diversity and density fluctuations in spatially growing populations.

There are several interesting directions for future research. 
Our framework can be easily extended to other reaction-diffusion systems, including two-dimensional and heterogeneous landscapes, to study how the interplay between different noise types shapes biodiversity patterns. It still remains to be explored how diffusive noise influences fixation times in the stepping stone model as well as in general phase-separating populations. It would also be interesting to study the differences between quenched (fixed) and annealed (random) initial conditions~\cite{derrida2009current2,krapivsky2012fluctuations,cividini2017tagged,banerjee2022role,krapivsky2014large} in such models of population genetics, which can be addressed within the MFT framework we have developed. Finally, our results suggest that the stochastic Fisher equation may have limitations in its applicability to systems characterized by strong migration rates. In the absence of noise, the nonlinear equation~\eqref{sde_allele} admits Fisher waves~\cite{fisher1937wave,kolmogorov2019studies} that propagate at a unique speed dependent on initial conditions~\cite{van1988front}; small demographic noise preserves the wave nature but corrects the front speed~\cite{brunet1997shift} in a noise-strength-dependent manner~\cite{doering2003interacting,hallatschek2009fisher}. Since diffusive noise could add further corrections, understanding how such Fisher wavefronts and genetic domain walls evolve under strong diffusive noise could provide new insights on the roughness and stability of interfaces in biological and chemical systems~\cite{hallatschek2010life,pechenik1999interfacial,lemarchand1996fractal}.

{\it Acknowledgments:} We thank R. Benzi, S. Chakraborty, and S. Prakash for discussions.  
\bibliography{Reference}

\newpage
\clearpage

\begin{widetext}

\section*{\large Supplemental Material for\\ ``Diffusive noise controls early stages of genetic demixing''}

This document provides supplemental figures and details related to the results presented in the main text.

\maketitle


\section{Details of numerical solution}
\subsection{Numerical integration of the heterozygosity  equation}
\label{sec_integration}
We numerically integrate the modified equation for heterozygosity, given in Eq.~\eqref{eq:Hetero_new} in the main text, repeated below
\begin{equation}
\partial_t H = 2D_d \partial_x^2 H - \frac{2D_r a}{N} H \delta(x) + \frac{2D_d a}{N} \partial_x^2 [H \delta(x)],
\label{eq:Hetero_new_supp}
\end{equation}
In order to numerically integrate the above equation, we use a finite difference scheme. For the time integration, we use the forward Euler method. Spatial derivatives involving the Laplacian are discretized using a second-order central difference scheme. A particular challenge arises from the third term in the RHS of Eq.~\eqref{eq:Hetero_new_supp}, which involves the second derivative of the Dirac delta function. To handle this, we regularize the Dirac delta by approximating it with a Gaussian function defined as
\begin{eqnarray}
 \delta_{\mu}(x) = \lim_{\mu \to \infty} \sqrt{\frac{\mu}{\pi}} \exp{(-\mu x^2)}.  
 \label{eq_mu}
\end{eqnarray}
For the central difference scheme, the standard deviation of the Gaussian, $1/\sqrt{2\mu}$, must be at least twice the spatial discretization \(\Delta x\).  As \(\mu\) increases, \(\Delta x\) must be refined accordingly. The time step \(\Delta t\) is chosen such that \(\Delta t = (\Delta x)^2\) to ensure stability of the numerical scheme. With these, Eq.~\eqref{eq:Hetero_new_supp} is approximated by
\begin{eqnarray}
H(x,t+\Delta t)&=&H(x,t)+\frac{2 D_d\Delta t}{\Delta x^2}\Big(H(x+\Delta x,t)+H(x-\Delta x,t)-2H(x,t)\Big)
 \nonumber\\&-&\frac{2   D_r a \Delta t}{N} H(x,t)\delta(x)
+\frac{2D_d a \Delta t}{N \Delta x^2}\Big(H(x+\Delta x,t) \delta(x+\Delta x) \nonumber\\&+&H(x-\Delta x,t) \delta(x-\Delta x)-2H(x,t) \delta(x)\Big).
\label{eq:discrete_H_x_t}
\end{eqnarray}
Once the solution for $H(x,t)$ is obtained, it is averaged to filter out fluctuations below the lattice length scale $a$. The coarse-grained field $H(\bar{x}, t)$ is defined as
\begin{equation}
H(\bar{x}, t) = \frac{1}{a} \sum_{|x - j \Delta x| < a} H(\bar{x} - j \Delta x, t) \, \Delta x,
\label{H_averaged}
\end{equation}
where $\bar{x} = i a$ and $i \in (-L/2a,\; L/2a]$. Note that \(a\) is the spacing between the lattice sites in the Monte Carlo simulation, whereas \(\Delta x\) denotes the spatial discretization used in the numerical scheme. The solution given in Eq.~\eqref{H_averaged} agrees well with the Monte Carlo simulations (as evidenced by Fig.~\ref{fig:hxt} in the main text).

\subsection{Scaling solution}
\label{subsec_scaling}
A scaling solution of Eq.~\eqref{eq:Hetero_new_supp} can be obtained when the spatial and temporal coordinates, as well as the parameters of the system, are appropriately scaled. When the spatial and temporal coordinates are rescaled as 
$x' \rightarrow x/L$ and $t' \rightarrow D_d t / L^2$, Eq.~\eqref{eq:Hetero_new_supp} remains invariant 
provided that the genetic drift constant $D_r$ and the parameter $\mu$ of the Gaussian function 
are also scaled as $D_d' \rightarrow \frac{D_r}{D_d} L^2$ and $\mu' \rightarrow \mu L^2$. Under this transformation, the equation takes the form
\begin{eqnarray}
\frac{\partial H(x',t')}{\partial t'}&=&2 \frac{\partial^2 H(x',t')}{\partial x'^2}-A~  H(x',t')\delta(x')+B~ \frac{\partial^2 H(x',t')\delta(x')}{\partial x'^2},
\label{eq:scaled_equation}
\end{eqnarray}
where $A=2 \frac{L}{a}  \frac{R_r}{R_d}\frac{1}{N}$ and $B=\frac{2}{N}\frac{a}{L}$ are dimensionless constants. Therefore, for a fixed value of $A$ and $B$, different solutions of Eq.~\eqref{eq:Hetero_new_supp} corresponding to different choices of system parameters exhibit a scaling collapse. We demonstrate this scaling behavior in Fig.~\ref{fig:subfig0}.

\begin{figure}[h]
    \centering      
    \includegraphics[width=0.9\textwidth]{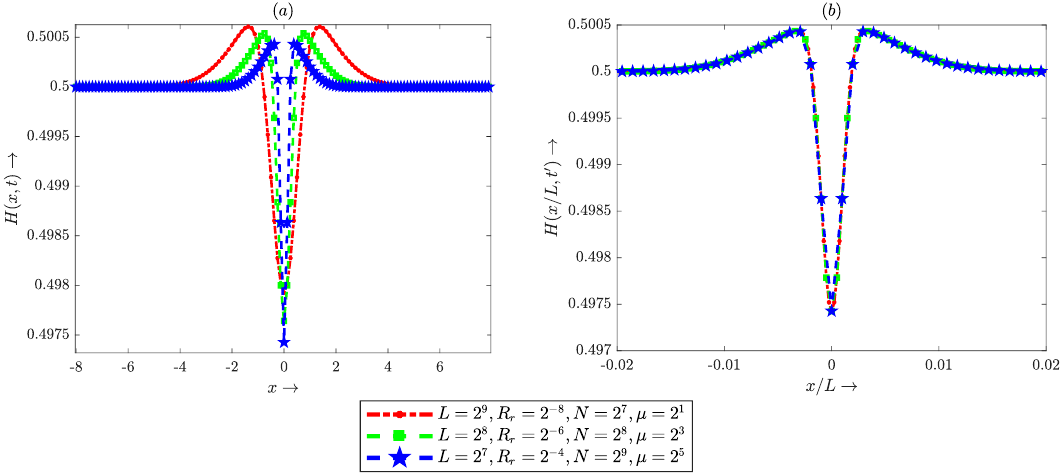}
    \caption{(a) Solution of Eq.~\eqref{eq:Hetero_new_supp} obtained using a finite difference method (Eq.~\eqref{eq:discrete_H_x_t}) with a spatial step size of $\Delta x = 1/2^3$ and a temporal step size of $\Delta t = 1/2^6$. Simulations are performed for different values of system parameters, $R_r$, $\mu$, and $L$ while keeping the dimensionless constants  $A=2^{-5} $ and $B= 2^{-15}$ fixed. The time instant for all the plots is fixed to be $t=2^6$.~(b)~When the spatial and temporal coordinates are rescaled as 
$x' \rightarrow x/L$ and $t' \rightarrow D_d t / L^2$, with the scaled time $t'=2^{-17}$ all the curves collapse onto a single universal curve. The diffusion rate for all the plots is taken to be $R_d=1$.
} 
    \label{fig:subfig0}
\end{figure}

\subsection{Late-time behavior of heterozygosity at the origin}
As discussed in the main text, the full dynamical behavior of the heterozygosity is accurately captured by Eq.~\eqref{eq:Hetero_new_supp}. In the long time limit, the effects of diffusive noise are less relevant as seen in Fig.~\ref{fig:hxt}(d). In this limit, Eq.~\eqref{eq:Hetero_new_supp} can be approximated as~\cite{korolev2010genetic}
\begin{equation}
\partial_t H = 2D_d \partial_x^2 H - \frac{2D_r a}{N} H \delta(x).
\label{eq:Hetero_old}
\end{equation}
The exact solution of Eq.~\eqref{eq:Hetero_old} has been derived previously~\cite{korolev2010genetic} and is given by
\begin{equation}
H(0,t)=H_0\mathrm{erfc}\left(\sqrt{\frac{t}{t_c}}\right)\mathrm{e}^{t/t_c},
\label{eq:theory_pred_supp}
\end{equation}
where $t_c=4D_d/(D_ra)^2 = 4D_dN^2/(R_ra)^2$ is the characteristic timescale. 
This solution predicts that for $t \gg t_c$: $H(0,t) \sim (t/t_c)^{-1/2}e^{-t/t_c}$. Therefore, when the time is scaled by $R_r^2$, an asymptotic power-law decay with exponent $-1/2$ is observed. This asymptotic behaviour is displayed in Fig.~\ref{fig:hetero_scaling_supp}(a). However, in finite systems of size $L$, this behaviour is cutoff at a finite time, as a crossover occurs when the diffusion length $\sqrt{D_d t}$ exceeds $L$. In this case a data collapse is achieved when time is rescaled by $R_r$ rather than $R_r^2$, which reflects the well-mixed (zero-dimensional) limit approached by small systems at late times, as shown from numerical solutions and Monte Carlo simulations in Fig.~\ref{fig:hxt}(d) in the main text, as well as from numerical solutions in Fig.~\ref{fig:hetero_scaling_supp}(b).

\begin{figure}[h]
    \centering
    \includegraphics[width=0.7\linewidth]{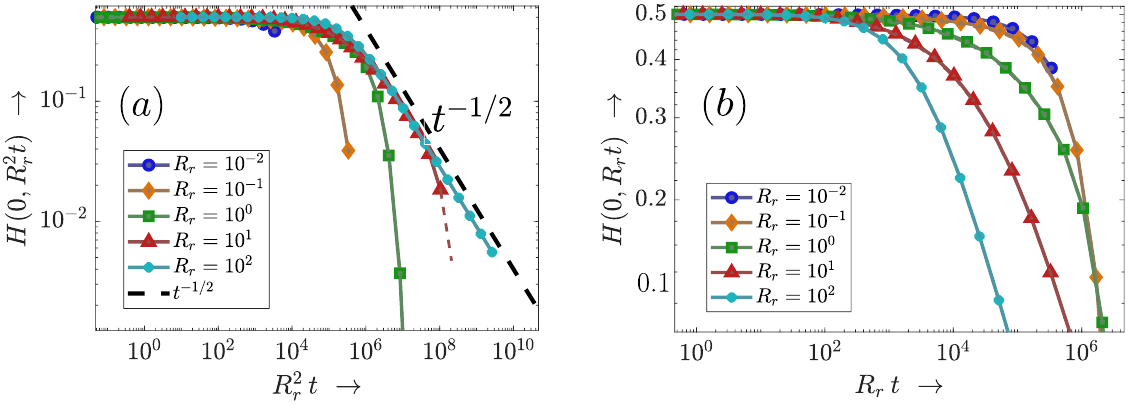}
    \caption{Numerical solutions of heterozygosity at the origin plotted for various reaction rates. (a) shows the scaling behaviour with the time axis scaled by $R_r^2$. The curves for larger reaction rates ($R_r = 10^1, 10^2$) collapse onto a single curve that displays a power-law decay ($t^{-1/2}$) over several decades as predicted in Eq.~\eqref{eq:theory_pred_supp} before finite-size effects become significant.  
     In ~(b), the time axis is scaled by $R_r$ in order to show the behaviour beyond this finite-size timescale. The curves corresponding to smaller reaction rates ($R_r = 10^{-2},\, 10^{-1},\, 10^0$) converge to the same asymptotic limit at late times, exhibiting behavior characteristic of a zero-dimensional system. For all the plots, the system size is taken to be $L=2^8$. The Dirac-Delta function is regularized with $\mu=6$. The space discretization for reaction rates $R_r=10^{-2},R_r=10^{-1},R_r=10^{0}$ is taken to be $ \Delta x=1/2^3$, whereas for larger reaction rates $R_r=10^{1},R_r=10^{2}$ it is $\Delta x=1/2^5$. 
     The time discretization is $\Delta t=\Delta x^2$ for all plots.
    }
    \label{fig:hetero_scaling_supp}
\end{figure}


\section{Details of microscopic simulations}
\subsection{Monte Carlo simulation parameters}

For the microscopic simulations, we use a kinetic Monte Carlo scheme~\cite{prados1997dynamical,bortz1975new,voter2007introduction}. We discretize the spatial domain into $L$ lattice sites, separated by a lattice constant $a$. Each site initially contains $N/2$ particles, each of type A and type B. The dynamics involve two processes: diffusion and reaction, which occur with rates $R_d$ and $R_r$, respectively. The simulation time step $\delta t$ is chosen such that $(R_d + R_r)\delta t < 1$. The probabilities of a diffusion event, a reaction event, and no event occurring during a given time step are $R_d \delta t$, $R_r \delta t$, and $1 - (R_d + R_r)\delta t$, respectively.

During a microscopic step, a site is selected at random from the $L$ sites, followed by a stochastic selection of the event (diffusion or reaction) based on the associated probabilities. For a diffusion event, one particle from the chosen site is randomly selected and exchanged with a particle from a neighboring site. For a reaction event, two particles are randomly drawn from the same site; one undergoes reproduction, while the other is removed.
One Monte Carlo step consists of $L$ such updates, during which each site is expected to be updated once on average. We consider quenched uniform initial conditions for simplicity, with $f_i(0) = 0.5$ for all realizations.

\begin{figure}[ht!]
    \centering
      \includegraphics[width=0.8\textwidth]{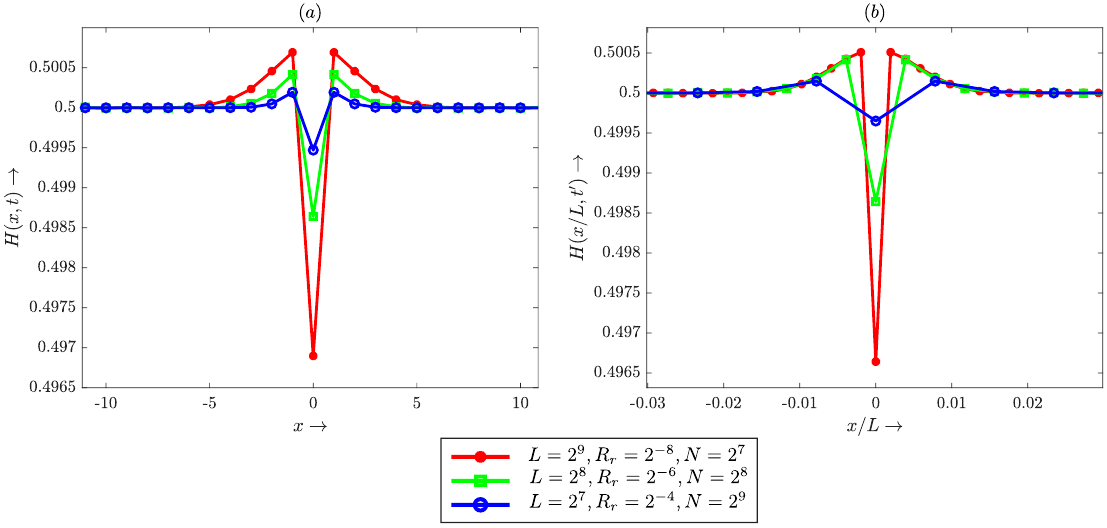}
        \caption{Scaling solutions obtained from Monte Carlo simulations. The heterozygosity profile $H(x,t)$ is plotted as a function of distance $x$.~(a) Multiple simulations are performed for different values of $L, N$, and $R_r$ so as to keep the dimensionless constants  $A= 2^{-5}$ and $B=2^{-15} $ in Eq.~\eqref{eq:scaled_equation} fixed. The time instant corresponds to all the plots are taken to be $t=2^7$. (b)~When the spatial and temporal coordinates are rescaled as 
$x' \rightarrow x/L$ and $t' \rightarrow D_d t / L^2$, with the scaled time $t'=2^{-17}$ all the curves collapse onto a single universal curve, verifying the scaling solution obtained numerically in Fig.~\ref{fig:subfig0}. The diffusive rate $R_d=1$ is fixed for all the plots.  }
        \label{fig:subfig1}
\end{figure}

\subsection{Parameters used for figures in the main text}
In this Section, we provide numerical values of various parameters used in the figures provided in the main text. In Fig.~\ref{fig: SS_Model} in the main text, the reaction rates are $K_1=K_2 = 1$ and the diffusive rate is $R_d=0.5$.  The system size is taken to be $L=256$, the Monte-Carlo time step is taken to be $\delta t=0.5$.
In Fig.~\ref{fig:hxt}(a) and Fig.~\ref{fig:hxt}(b) in the main text, the reaction rate  \( R_r = 0.1 \) and diffusion rate \( R_d = 0.5 \). The system size is $L=2^6$. To solve Eq.~\eqref{eq:Hetero_new} in the main text, we use a finite difference scheme (explained in Sec.~\ref{sec_integration}), with the spatial and temporal discretization $\Delta x=1/2^7$ and $\Delta t=1/2^{14}$ respectively. The Dirac delta function is regularized with $\mu=24$. In Fig.~\ref{fig:hxt}(d) in the main text, $\Delta x=1/2^3$, $\Delta t=1/2^{6}$ and $\mu=6$. In Fig.~\ref{fig:Q2_Dg} in the main text, the scaled reaction rate is taken to be $\tilde{R_r}=1$, diffusion rate $R_d=1$. For all figures in the main text, the lattice spacing used in the Monte Carlo simulations is \(a = 1\), and the number of particles per site (deme size) is \(N = 100\).

\subsection{Scaling solution using Monte Carlo simulations}
As discussed in Sec.~\ref{subsec_scaling},
Eq.~\eqref{eq:Hetero_new_supp} displays a scaling solution when space and time are rescaled as follows: $x' \rightarrow x/L$, $t' \rightarrow D_st/L^2$, provided $A=2 \frac{L}{a} \frac{R_r}{R_d}\frac{1}{N}$ and $B=\frac{2}{N}\frac{a}{L}$ are held fixed. This scaling behaviour is confirmed using direct Monte Carlo simulations in Fig.~\ref{fig:subfig1}.

\section{Convergence tests}
The numerical solution of Eq.~\eqref{eq:Hetero_new_supp} obtained from the integration scheme in Eq.~\eqref{eq:discrete_H_x_t} converges to an asymptotic solution in the limit as $\mu \rightarrow \infty$, $\Delta x \rightarrow 0$, and $\Delta t \rightarrow 0$. 
In the following, we validate our numerical scheme for different times $t$ across various parameters: lattice spacing $a$ used in the Monte Carlo simulations, deme size $N$, as well as the sharpness of the delta function parametrized by $\mu$ in Eq.~\eqref{eq_mu}. Fig.~\ref{fig:H_mu_t_a_1} demonstrates the stability of the scheme for different values of $\mu$. The numerical results are in good agreement with Monte Carlo simulations at all times shown in panels (a) to (d). A similar analysis is presented in Fig.~\ref{fig:H_mu_t_a_0.5}, where the lattice spacing is set to $a = 0.5$, confirming that $H(x,t)$ also depends on the value of $a$ as is clear from Eq.~\eqref{eq:Hetero_new_supp}. Finally, in Fig.~\ref{fig:H_delx_t_a_0.5}, we confirm the stability of the scheme across various spatial discretizations $\Delta x$ used in the numerical integration. 

\begin{figure}[h]
    \centering
    \includegraphics[width=1\linewidth]{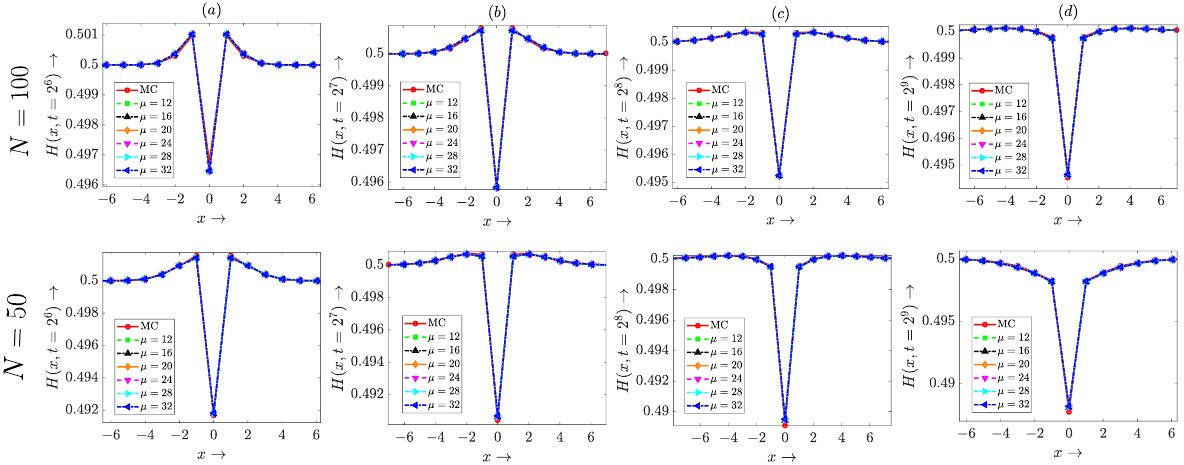}
    \caption{The heterozygosity profile $H(x,t)$ is plotted as a function of distance $x$ for different values of $\mu$, showing good convergence as well as agreement with the Monte Carlo simulations (MC). The lattice spacing in the Monte Carlo simulations is set to unity ($a=1$). The diffusion rate $R_d = 0.5$, reaction rate $R_r = 0.1$, and time step $\delta t = 1$. The system size is $L = 2^6$. For numerical integration, the spatial discretization is chosen as $\Delta x = \frac{1}{2^7}$ and the time discretization as $\Delta t = \Delta x^2$. The top panel corresponds to deme size $N=100$ and the bottom panel corresponds to $N=50$.}
    \label{fig:H_mu_t_a_1}
\end{figure}
\begin{figure}[h]
    \centering
    \includegraphics[width=1\linewidth]{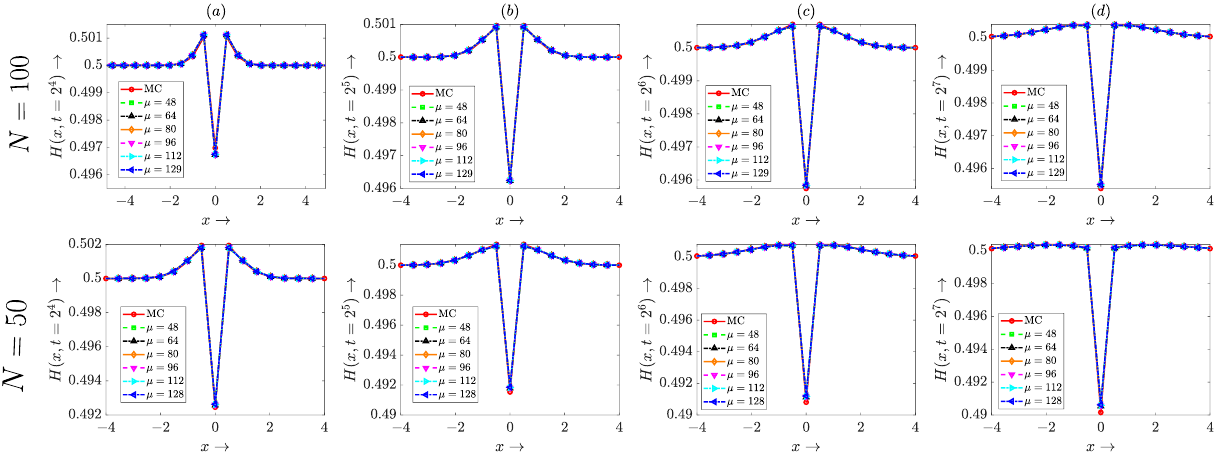}
    \caption{The heterozygosity profile $H(x,t)$ is plotted as a function of distance $x$ for different values of $\mu$, showing good convergence as well as agreement with the Monte Carlo simulations (MC). In the Monte Carlo simulations, the lattice spacing is set to $a=1/2$.  The diffusion rate $R_d = 0.5$, reaction rate $R_r = 0.1$, and time step $\delta t = 1/2^2$. The system size is $L = 2^6$. For the finite difference method, the spatial discretization is chosen as $\Delta x = \frac{1}{2^7}$ and the time discretization as $\Delta t = \Delta x^2$. The top panel corresponds to deme size $N=100$, and the bottom panel corresponds to $N=50$.}
    \label{fig:H_mu_t_a_0.5}
\end{figure}
\begin{figure}
    \centering
    \includegraphics[width=1\linewidth]{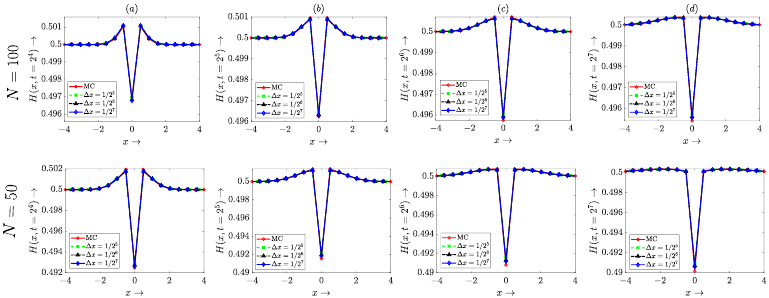}
    \caption{The heterozygosity profile $H(x,t)$ is plotted as a function of distance $x$ for different values of $\Delta x$, showing good convergence as well as agreement with the Monte Carlo simulations (MC). In the Monte Carlo simulation, the lattice spacing is set to $a=1/2$. The diffusion rate $R_d = 0.5$, reaction rate $R_r = 0.1$, and time step $\delta t = 1/2^2$. The system size is $L = 2^6$. The time discretization in the finite difference method is chosen as $\Delta t = \Delta x^2$. The parameter $\mu$ of the Gaussian function appearing in Eq.~\eqref{eq_mu} is set to $\mu=48$ for all the plots. The top panel corresponds to deme size $N=100$, and the bottom panel corresponds to $N=50$.}
    \label{fig:H_delx_t_a_0.5}
\end{figure}

\clearpage
\end{widetext}
\end{document}